\documentclass[aps,prb,twocolumn,showpacs,showkeys,preprintnumbers,amsmath,amssymb]{revtex4-2}

\usepackage[utf8]{inputenc}
\usepackage{graphicx}
\graphicspath{{Figures/}}
\usepackage{hyperref}
\usepackage{units}
\usepackage{amsmath,amssymb}
\usepackage{textgreek}
\usepackage{multirow}
\usepackage{rotating}
\usepackage{color}
\usepackage{ulem}

\begin{document}
\title{Intercalant-induced Kekulé ordering and gap opening in quasi–free-standing graphene}
\author{Huu Thoai Ngo$^{1}$} 
\author{Zamin Mamiyev$^{1}$} 
\author{Niklas Witt$^{2}$} 
\author{Tim Wehling$^{3,4}$}
\author{Christoph Tegenkamp$^{1}$} \email{christoph.tegenkamp@physik.tu-chemnitz.de}
\address{$^{1}$Institut f\"ur Physik, Technische Universit\"at Chemnitz, Reichenhainer Str. 70, D-09126 Chemnitz, Germany.}
\address{$^{2}$Institute for Theoretical Physics and Astrophysics, University of Würzburg, Am Hubland, 97074 Würzburg, Germany.}
\address{$^{3}$Institute of Theoretical Physics, University of Hamburg, Notkestrasse 9, 22607 Hamburg, Germany.}
\address{$^{4}$The Hamburg Centre for Ultrafast Imaging, Luruper Chaussee 149, 22761 Hamburg, Germany.}
\date{\today}
\begin{abstract}
We present a comprehensive investigation of the structural and electronic properties of Sn-intercalated buffer layers on SiC(0001) using low-temperature scanning tunneling microscopy and spectroscopy (LT-STM/STS), spot-profile analysis low-energy electron diffraction (SPA-LEED), and density functional theory (DFT) calculations. Sn intercalation effectively decouples the buffer layer, yielding quasi–free-standing monolayer graphene (QFMLG) while introducing local lattice distortions. Bias-dependent STM imaging revealed the coexistence of conventional and Kekulé-ordered graphene domains, governed by the underlying Sn(1×1) reconstruction at the SiC interface. The measured STS spectra exhibit good agreement with DFT results. However, achieving homogeneous Sn(1×1) domains remains challenging, apparently, due to strain within the Sn monolayer, which drives the emergence of Kekulé distortions and the associated electronic band-gap opening homogeneously in graphene.  These findings highlight the crucial role of intercalant homogeneity and strain in tuning graphene’s structural and electronic properties.  
\end{abstract}

{}

\maketitle

\section{Introduction}
Graphene stands out for its exceptional electronic and structural properties \cite{Geim2007}. Its peculiar crystal structure comprises two symmetric carbon sublattices that cause linear band dispersion, where the conduction and valence bands touch at the Dirac point \cite{Geim2007, Wehling2014}. This feature results in a zero-band gap, high carrier mobility, and electrical conductivity, making graphene a promising material for nanoelectronics \cite{Novoselov2004, Zhang2005, CastroNeto2009}. However, the intrinsic lack of a band gap limits the use of pristine graphene in optoelectronic, transistor, and spintronic devices. Consequently, tailoring graphene’s electronic properties has become a central research topic to enable broader applications.

Several strategies have been explored to induce a band gap in graphene, which can be classified according to the underlying mechanism: (i) quantum confinement in graphene nanoribbons \cite{Brey2006, Son2006, Nhung2023} and quantum dots \cite{Yan2018}, (ii) breaking of the sublattice symmetry by interfaces  \cite{Varykhalov2010, Zhou2007, Ghosal2022}, chemical functionalization of graphene \cite{Balog2010, Son2016}, intercalation \cite{Bao2021}, or strain engineering \cite{Yan2013} (iii) stacking of multiple graphene layers \cite{AOKI2007123, Zhang2009}, and (iv) electronic proximity effects via spin-orbit coupling (SOC) from heavy-elements or electronically correlated systems \cite{Jeon2013, Warmuth2016, Cano2025, Ghosal2025}. 

In this context, lattice distortions in graphene have been shown to give rise to novel electronic phases, most notably Kekul\'e graphene \cite{Blason2022}. This structurally modified phase of graphene is characterized by the formation of a $(\sqrt{3} \times \sqrt{3})R30^{\circ}$ superlattice, which folds the Dirac cones from K and K
’ points into the Γ point of the Brillouin zone \cite{Bao2021, Eom2020, Mamiyev2025}. The resulting band structure leads to remarkable electronic features, including a gap opening \cite{Bao2021, Eom2020, Qu2025}, charge fractionalization \cite{Bergman2013, Guan2024}, and valley-momentum locking \cite{Gamayun2018} depending on coupling strength of the K and K' valleys. 
Thereby, two distinct types of Kekul\'e ordering were identified. The Kekulé-Y phase, observed in epitaxial graphene grown on Cu substrates \cite{Gutierrez2016}, and hydrogen-adatom-absorbed graphene/SiC \cite{Guan2024} still preserves the gapless nature as pristine graphene. By contrast, the Kekul\'e-O phase involves a three-fold symmetric distortion of the graphene honeycomb lattice (“O”-shape pattern), which opens an energy gap at the Γ point \cite{Qu2025}. 

To date, the emergence of Kekulé phases in graphene has not been systematically explored.  Epitaxial graphene growth on the Si-terminated surface of silicon carbide (SiC) is a promising approach for producing scalable, high-quality graphene layers which can be further functionalized by adsorption and intercalation \cite{Emtsev@2008, Nair@2017}.  Pronounced local perturbations of the carbon bonding environment, for instance through lithium adsorption and intercalation, can induce effective sublattice coupling \cite{Bao@2022, Bao2021, Qu2022}. In contrast, for homogeneously intercalated systems the situation remains ambiguous: although many exhibit well-ordered periodic intercalant structures, including hydrogen-intercalated QFMLG, there have been no prior reports of Kekulé-type distortions. For instance, Pb-intercalated quasi–free-standing monolayer graphene (QFMLG) exhibits a moiré superstructure accompanied by a small band-gap opening at the K point, which is primarily attributed to a symmetry breaking between the graphene sublattices\cite{Ghosal2022, Gruschwitz@2025}.
In case of Sn, Mamiyev et al. reported more recently the emergence of Kekul\'e-O graphene in the Sn intercalated buffer layer \cite{Mamiyev2025}. However, the atomic structure and details of the electronic properties of the Kekul\'e-O graphene in the heavy-element intercalated buffer layer remained fully unexplored. 

In this work, we intercalate Sn into the buffer layer/SiC interface under ultrahigh vacuum (UHV) conditions. The structural and electronic properties of the Sn-intercalated buffer layer on SiC was comprehensively investigated using low-temperature scanning tunneling microscopy/spectroscopy (LT-STM/STS), supplemented by spot-profile analysis, low-energy electron diffraction (SPA-LEED), and density functional theory (DFT) calculations. SPA-LEED and LT-STM are used to monitor the intercalation process and to identify Sn-intercalated phases.  Our STM results clearly show that a monolayer of intercalated Sn gives rise to two distinct electronic phases: Kekul\'e-modulated graphene with an energy gap and metallic, charge-neutral pristine graphene. Our findings provide insight into how Sn intercalants modify the graphene/SiC interface and the local electronic band structure of graphene at both micro- and nanoscales. 


\section{Methods and experimental details}

Buffer layers (BL) on SiC  were epitaxially grown by thermal sublimation on the Si-terminated surface of a 4H-SiC(0001) substrate, followed by adsorption of Sn and subsequent annealing for intercalation under UHV conditions and elevated temperatures. For further details the reader is referred to Refs. ~\cite{Emtsev2009, Mamiyev2022, Mamiyev2024, Mamiyev2025}. The intercalation process was controlled and monitored by SPA-LEED.  Afterward, the sample was transferred via a UHV suitcase into the STM chamber, where its structural and electronic properties were investigated at cryogenic temperatures (T = 77 K, 4.5 K) under UHV conditions. The STM tips used were made from PtIr wires, which were calibrated on an Au (111) substrate. The STM measurements were performed in constant current mode. Tunneling spectroscopy (STS) was conducted using an SR830 DSP lock-in amplifier with a modulation frequency of 860 Hz and modulation voltages of V$_{rms}$ = 7 mV and 50 mV. 

We studied the electronic structure of the graphene/Sn/SiC heterostructure using density functional theory (DFT) as implemented in the Vienna ab initio Simulation Package (VASP) \cite{Kresse@1993, Kresse@1996, G.Kresse@1996}. We employed the Perdew--Burke--Ernzerhof generalized gradient approximation \cite{Perdew@1996} together with projector augmented-wave basis sets \cite{Blochl@1994, Kresse@1999} and a plane-wave cutoff energy of 400 eV. Long-range dispersion effects were included via DFT-D3 van der Waals corrections \cite{Grimme@2010}.

We relaxed the heterostructure using a conjugate-gradient scheme until all residual forces were below 0.005 eV \AA$^{-1}$ and applied a vacuum region of approximately 12~\AA ~ to avoid spurious interactions between periodic images. Self-consistent calculations were performed on a $30 \times 30 \times 1$ $k$-mesh, while densities of states (DOS) were evaluated using a denser $60 \times 60 \times 1$ mesh and the tetrahedron method with Bl\"ochl corrections \cite{E.Bloch@1994}.

To capture the key experimental features at manageable computational cost, we adopted an approximate $\sqrt{3} \times \sqrt{3}$ surface supercell, with graphene commensurately strained into a $2 \times 2$ arrangement relative to the substrate, following the approach commonly used for graphene/SiC systems \cite{Mattausch@2007, Federl2025, Ghosal2025}. The spectrum obtained from theory was shifted so that the Dirac point coincides with the charge-neutrality measured by ARPES and STS \cite{Federl2025}. The DFT data underlying this work are available at~[XX]. 

\section{Results and Discussions}
\begin{figure*}[tb]
	\centering
	\includegraphics[width=0.65\linewidth]{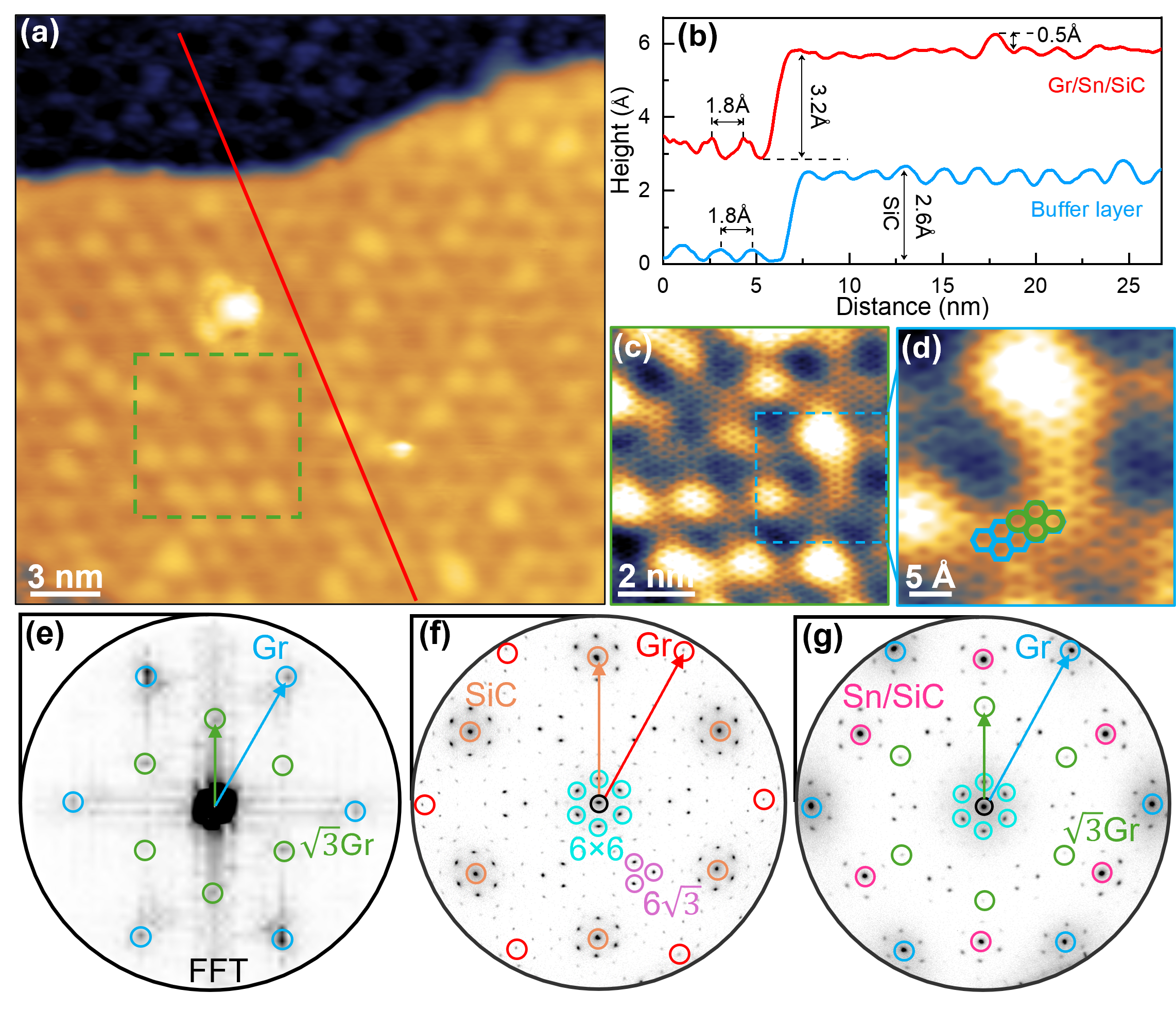}
	\caption{(a) STM topographic image of a large Sn-intercalated area, showing bubble-like features (+1.5 V, 250 pA). (b) Height profile extracted along the red line in Fig.~\ref{Figure1}(a), in comparison to the height profile across a SiC step with buffer layer (blue). (c) Atomically resolved STM image of a 7 × 7 $\rm nm^2$ area outlined by the green dashed rectangle in Fig.~\ref{Figure1}(a) (+1.2 V, 300 pA). (d) A zoomed-in STM image of the area marked by the blue dashed rectangle (Fig.~\ref{Figure1}(c)), exhibiting the graphene (blue hexagon) and O-shape Kekul\'e structure (green circle).  (e) Corresponding FFT image of Fig.~\ref{Figure1}(c), showing the diffraction spots of the graphene (blue circles, $a = 2.46$ \AA) and Kekul\'e-O (green circles, $a = 4.25$ \AA). (f-g) SPA-LEED images of the clean BL (f) and after the Sn intercalation (g), recorded with 100 eV primary energy at 300~K. The pink circles represent the diffraction spots of the intercalated Sn(1$\times$1) phase. All STM images were acquired at 77 K.  
\label{Figure1}} 
\end{figure*}

After room-temperature Sn deposition and subsequent annealing at 1120 K,  the sample homogeneity was examined on larger scales by means of  LT-STM. This intercalation procedure has previously been shown to yield a Sn-monolayer  interface structure with a long-range-ordered graphene sheet on top \cite{Mamiyev2024, Harling2025}. The STM image shown in Fig.~\ref{Figure1}(a) reveals a  Sn-intercalated surface with an apparent height of 3.2~\AA\ with respect to the buffer layer (BL), corresponding to quasi-free monolayer graphene (QFMLG) (Fig.~\ref{Figure1}(b), red curve). To clarify this height value, we extracted a height profile across a SiC step hosting only BL, as shown in Fig.~\ref{Figure1}(b) (blue curve). The measured SiC step height is 2.6~\AA\ \cite{Pakdehi2020}, while the spacing between two nearest peaks is approximately 1.8~\AA\, corresponding to the SiC(6$\times$6) lattice constant \cite{TOK2004145}. Additionally, we also observed bubble-like features with various shapes and sizes, similar to the phase reported recently (see SI, Fig.~\ref{Figure1}(e)\cite{Mamiyev2025}. Despite the qualitative agreement,  the measured height of a bubble is 0.5~\AA\ relative to QFMLG is slightly smaller, as well as the lateral size and distribution, which is apparently due to a longer annealing time.  

The atomically resolved STM images, shown in Figs.~\ref{Figure1}(c) and \ref{Figure1}(d),  display the graphene lattice along with the bubble-like features, indicating a Sn(1$\times$1) phase formed at the interface. In fact, the use of high temperature annealing (1120 K) in this study leads to full desorption of the deposited Sn atoms and forms slightly smaller clusters at the interface, compared to our previous study \cite{Mamiyev2025}. As shown in Figs.~\ref{Figure1}(a)-(c) and the corresponding line scan (see also SI, Figs.~\ref{Figure1}(b)-(d)), the Sn clusters randomly occupy positions on the SiC(6$\times$6) supercell grid, similar to the case of Eu-intercalated graphene on SiC \cite{Anderson@2017}. Apparently, the former bonds of the BL to SiC  trigger the random arrangement of Sn intercalants on the SiC substrate and are in line with our previous insights on the Sn intercalation forming Mott phases \cite{Ghosal2025}. Furthermore, close inspection of the high resolution STM images revealed the coexistence of two distinct graphene lattices, including the conventional graphene lattice and the Kekul\'e-modulated graphene (Fig.~\ref{Figure1}(d)). This finding is confirmed by the corresponding FFT image, where $(\sqrt{3} \times \sqrt{3})R30^{\circ}$ spots appear with respect to the graphene lattice (Fig.~\ref{Figure1}(e)) and is in good agreement with the SPA-LEED results. 

Fig.~\ref{Figure1}(f) presents a high-resolution SPA-LEED image of the BL on 4H-SiC(0001), acquired at 100 eV primary energy. The diffraction pattern exhibits sharp SiC(1$\times$1) and ZLG(1$\times$1) spots, along with distinct spots corresponding to the $(6\sqrt{3}\times6\sqrt{3})R30^{\circ}$ reconstruction. A 6$\times$6 quasi-periodicity is also observed, arising from the broken symmetry of the topmost SiC layer relative to the bulk.
In panel \ref{Figure1}(g) we show the diffraction pattern after Sn intercalation, indicating the transformation of the BL into freestanding graphene, e.g., evidenced by the brighter Gr(1$\times$1) diffraction spots and an intense, coherent background \cite{Mamiyev2022, Chen2019}. Also, the SPALEED pattern measured at room temperature reveal  $(\sqrt{3}\times\sqrt{3})R30^{\circ}$ spots with respect to the graphene lattice.

\begin{figure*}[tb]
	\centering
	\includegraphics[width=0.8\linewidth]{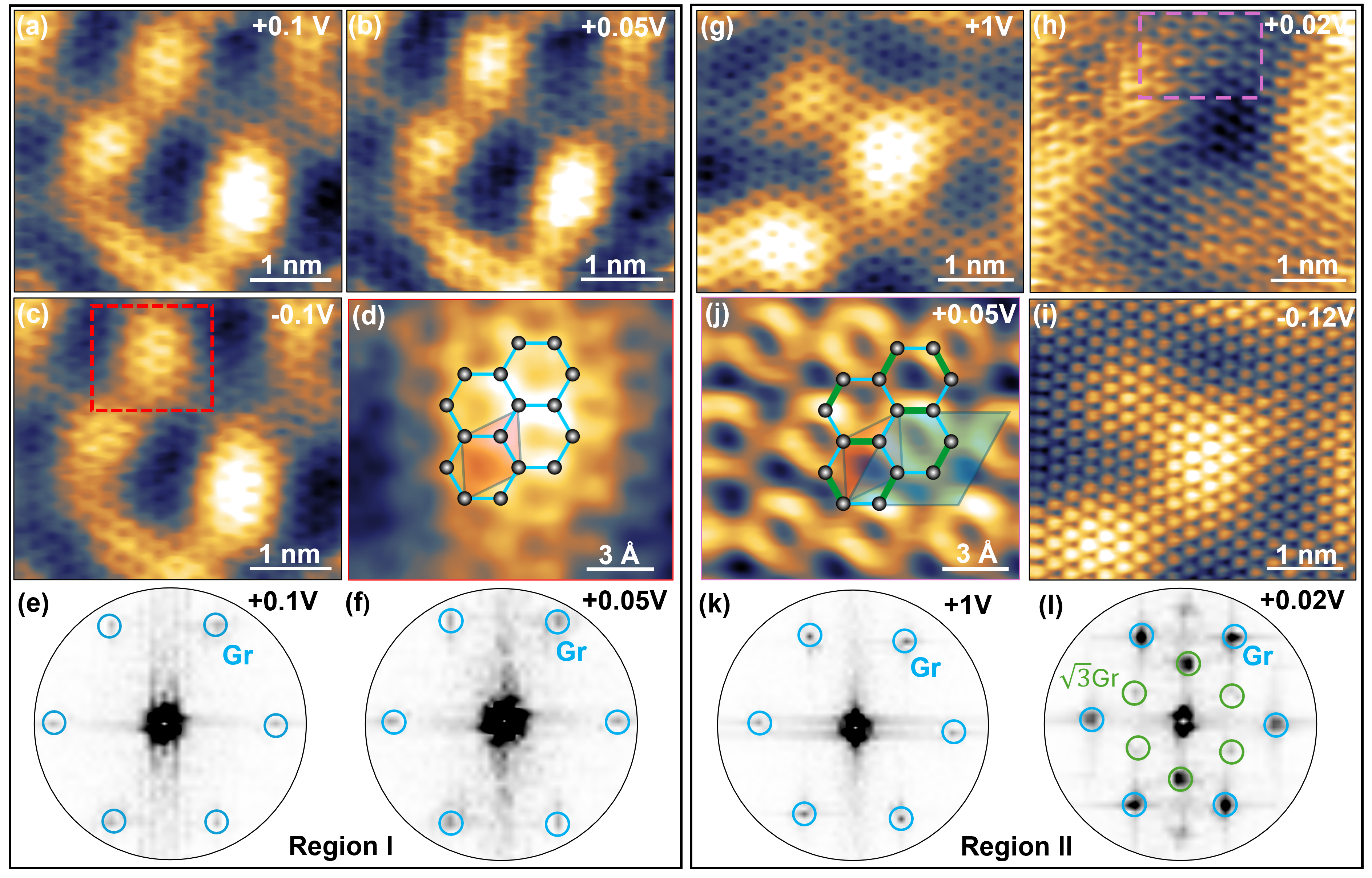}
	\caption{Bias-dependent STM images showing the topography of two different Sn-intercalated BL regions. (a-d) Region I: STM images of Sn-induced free-standing MLG. (a) +0.1 V - 380 pA, (b) +0.05 V - 360 pA, (c) -0.1 V - 370 pA. (d) A closed-up STM image of the area marked by the red dashed rectangle in Fig.~\ref{Figure2}(c), showing the graphene honeycomb lattice.  (e-f) Corresponding FFT images of Fig.~\ref{Figure2}(a) (e), and of Fig.~\ref{Figure2}(b) (f). (g-j) Region II: STM images of Sn-induced distorted graphene structure. (g) +1 V - 300 pA, (h) +0.02 V - 280 pA, (i) -0.12 V - 300 pA. (j) Zoomed-in STM image of the area marked by the purple dashed rectangle in Fig.~\ref{Figure2}(h), showing the Kekul\'e-O graphene lattice (+0.05 V, 400 pA). (k-l) Corresponding FFT images of Fig.~\ref{Figure2}(g) (k) and of Fig.~\ref{Figure2}(h) (l). Blue and green circles represent the diffraction spots of graphene and Kekul\'e-graphene ($\sqrt{3}$Gr). Schematic representations of the graphene and Kekul\'e-O graphene lattices are also shown in Figs.~\ref{Figure2}(d)-(j). The green lines represent the distorted C-C bonds. All STM images were acquired at 4.5 K.
\label{Figure2}} 
\end{figure*}

In order to obtain further insights into the structural properties, we performed bias-dependent STM measurements for the  two characteristic Sn-intercalated regions. In region I (Figs.~\ref{Figure2}(a)-(d)),  the STM images show the graphene honeycomb structure superimposed by bubble-like features even at low biases, demonstrating the robust presence of the Sn(1$\times$1) phase at the buffer layer/SiC interface.  The corresponding FFT images display the hexagonal graphene lattice with a corresponding lattice parameter of 2.46 \AA\ (Figs.~\ref{Figure2}(e)-(f), blue circles). In region II (Figs.~\ref{Figure2}(g)-(j)), we also observed a hexagonal graphene lattice and bubble-like features associated with underlying Sn atoms at comparably high bias voltages (Fig.~\ref{Figure2}(g)), in good agreement with previous observations. However, in this region, Kekul\'e-distorted graphene emerges when the bias voltage is lowered to +0.02 V (Fig.~\ref{Figure2}(h)). This distorted graphene structure is clearly visible in (Fig.~\ref{Figure2}(j)), where the bond symmetry breaking manifests as an ordered arrangement of $(\sqrt{3}\times\sqrt{3})R30^{\circ}$ unit cell, as obvious from the FFT image (Fig.~\ref{Figure2}(l), green circles).

\begin{figure*}[tb]
	\centering
	\includegraphics[width=0.8\linewidth]{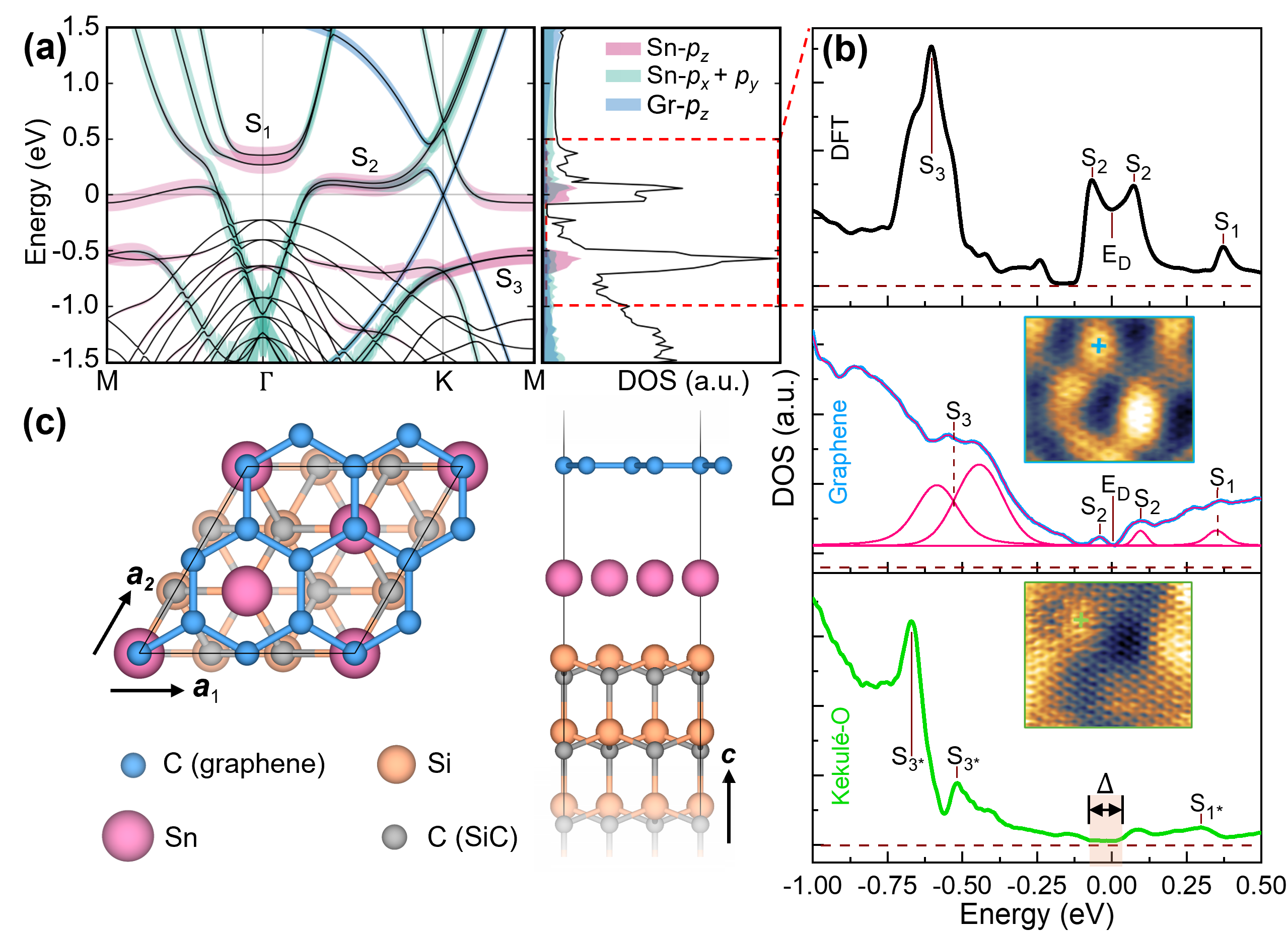}
	\caption{(a) DFT calculations of the band structure (left) and density of states (right) for Sn-intercalated BL/SiC. (b) Comparison between the calculated (DFT) and experimental (STS) electronic structure of EG/Sn(1$\times$1)/SiC. (Top) Calculated DOS within an energy range (-1 eV to +0.5 eV) extracted from Fig.~\ref{Figure3}(a). (Middle) STS of the Sn-induced QFMLG recorded at $V_b$ = 1 V, $I_t$ = 350 pA (blue curve). The spectrum was fitted with a Gaussian function. (Bottom) STS of the Sn-induced Kekul\'e-O graphene recorded at $V_b$ = 1 V, $I_t$ = 400 pA (green curve).  $S_1*$ and $S_3*$ represent the modification of the Sn-electronic bands. The brown dashed lines represent the zero DOS. The inserted STM images display the probed positions. (c) Structural model for EG/Sn(1$\times$1)/SiC with $(\sqrt{3} \times \sqrt{3})$ supercell: top view (left) and side view (right). The STS data were recorded at 4.5 K.
\label{Figure3}} 
\end{figure*}

In the following, we will discuss the electronic structure measured in both regions. In order to better understand the influence of Sn intercalants on the electronic properties measured with STS, density functional theory (DFT) calculations were carried out for Sn(1$\times$1)-intercalated buffer layer on SiC (0001). As shown in Fig.~\ref{Figure3}(a) (left panel), the graphene-derived $\pi$ bands (blue) retain their linear dispersion near the Fermi energy level, confirming the effective decoupling of the buffer layer after Sn intercalation. The apparently small overlap between the Sn $p_{x,y}$- and $p_z$-orbitals results in rather flat bands, giving rise to three characteristic orbital-projected DOS (pDOS) features, termed $S_1$, $S_2$, and $S_3$. The $S_3$-band has the largest spectral weight, about 0.5~eV below the Dirac point of graphene. A strong hybridization with the $\pi$-band of graphene is seen only for the $S_2$ state in the vicinity of the Dirac point, while the other Sn bands show almost no hybridization with graphene.

In Fig.~\ref{Figure3}(b), we compare the DFT results with STS measurements performed in the two regions. The spectrum shown in the middle panel can be reasonably explained by adopting the peak structure obtained from the DFT calculations, shown at the top, together with a linear increase in the DOS arising from the presence of graphene. Please note, the graphene layer sits atop the Sn layer, i.e., the electron transmission across the tunneling barrier from the tip into the graphene is more likely than a tunneling directly into the Sn layer. As a result, the graphene DOS contribution to the d$I$/d$V$ spectrum is not fully captured in the DFT calculations. Nevertheless, from the fit of the data, the ratio of the spectral weights is reproduced reasonably well within the expected limits (Fig.~\ref{Figure3}(b), blue curve). 

The formation of the Kekul\'e phase is clearly reflected in the opening of a band gap of about 90~meV (Fig.~\ref{Figure3}(b), green curve). In addition, the peaks that were previously assigned to the Sn states appear to undergo significant changes. The emergence of the Kekul\'e structure is accompanied by the back-folding of the graphene band structure to the $\Gamma$ point and the opening of a band gap, fully consistent with our observations. If we assume that significant amounts of low-energy graphene spectral weight are folded to the $\Gamma$ point and that there is significant hybridization between graphene and Sn, the spectral changes can be understood as follows:
Based on the calculations for the unreconstructed graphene phase, the back-folding of the graphene bands should be associated with strong modifications of the $S_1$ and $S_2$-Sn bands. Some spectral signatures of the original $S_3$ state, however, appear to participate only minimally in the new hybridization. In addition, STM probes states preliminary at the $\Gamma$ point, i.e., the spectral weight of the Sn states appears weaker compared to graphene-related states in the Kekul\'e phase.  A similar observation of the hybridization and coupling with the folded Dirac bands at the $\Gamma$ point has also been reported in Eu-intercalated graphene/SiC \cite{qiu@2025}. This study suggests the essential role of metal intercalants in stabilizing and modulating the electronic states of Kekul\'e-O graphene. Moreover, both d$I$/d$V$ spectrums show a finite DOS offset at zero bias, implying the metallic characteristic of the Sn(1$\times$1) phase, as deduced by the DFT calculation.

 \begin{figure*}[tb]
	\centering
	\includegraphics[width=0.9 \linewidth]{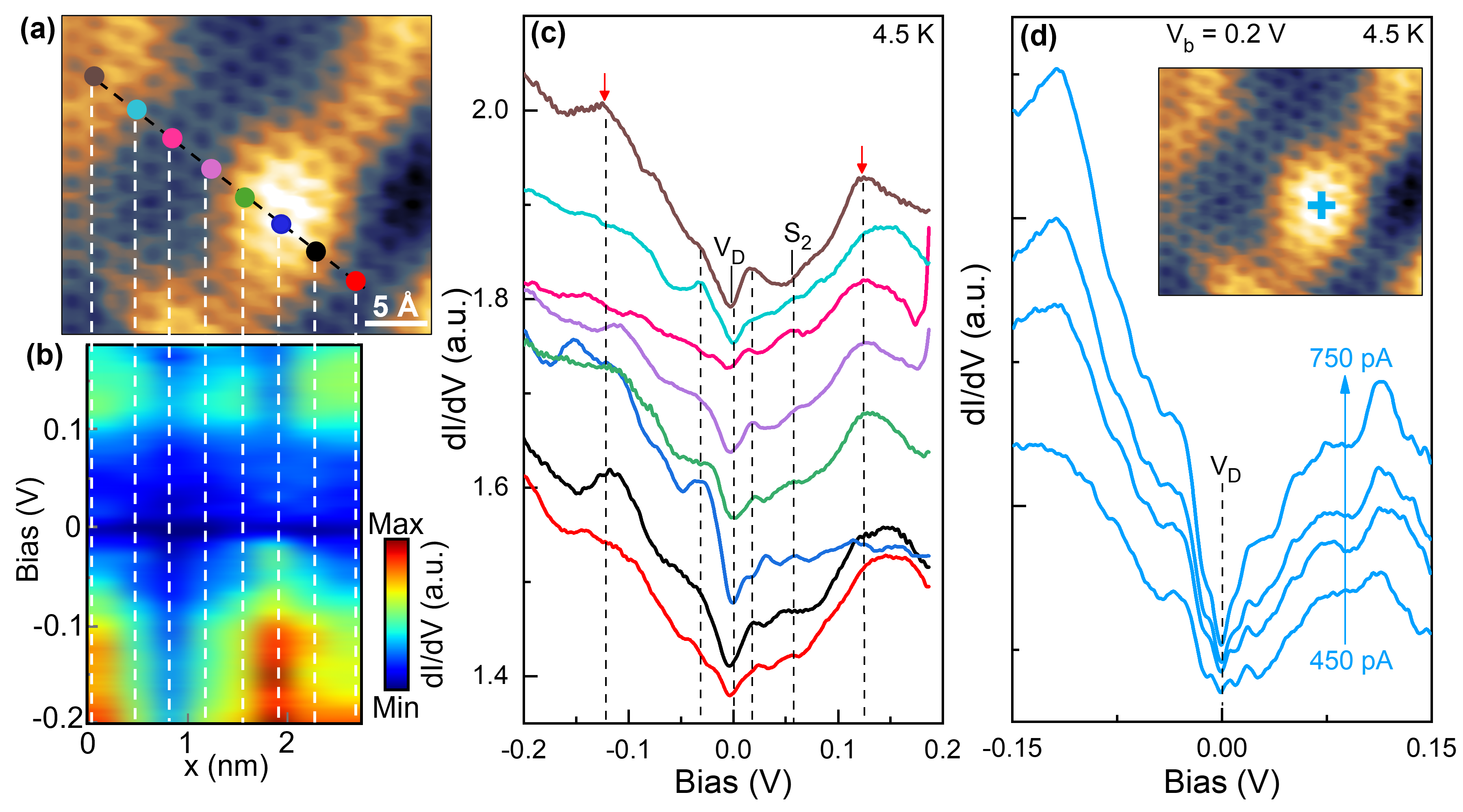}
	\caption{(a) Topographic STM image of a Sn-intercalated BL/SiC (region I) ($V_b$ = 0.1 V, $I_t$ = 380 pA). (b) Line scan of d$I$/d$V$ ($V_b$, x) spectra measured at different positions along the black dashed line in the STM image ($V_b$ = 0.2 V, $I_t$ = 400 pA). (c) Corresponding d$I$/d$V$ spectra (STS) extracted from the specific position as indicated by the colored circles in the STM image. (d) Set-point current-dependent d$I$/d$V$ spectra recorded at $V_b$ = 0.2 V, and $I_t$ = 450 pA, 550 pA, 650 pA, 750 pA. The inserted STM image shows the probed position. All STM/STS data were acquired at T = 4.5 K. 
\label{Figure4}} 
\end{figure*}

To further elucidate the influence of the inhomogeneity of the Sn intercalants, STS measurements were performed at various positions across the Sn-intercalated QFMLG regions,  indicated in Fig.~\ref{Figure4}(a). The acquired d$I$/d$V$ spectra exhibit the characteristic V-shaped profile with the Dirac point located at zero bias (Fig.~\ref{Figure4}(c)), consistent with charge-neutral graphene and in agreement with photoemission results \cite{Federl2025}. 
Notably, the acquired spectra exhibit a finite offset in the d$I$/d$V$ intensity at zero bias, demonstrating the formation of a metallic Sn(1$\times$1) phase at the interface, in agreement with DFT calculations and ARPES results \cite{Federl2025}.

The d$I$/d$V$ spectra show only minor variations in the peak intensities among the probed positions, and no shifts in the positions.  The two broad peaks at $\pm$0.12 eV in panel \ref{Figure4}(c) marked by red arrows, are related to the $S_2$ bands seen in Fig.~\ref{Figure3}(b)). The variation in the $S_2$ state intensity shown in panel \ref{Figure4}(c) likely reflects changes in the relative coupling between the graphene layer and the underlying Sn interface layer. In addition, the STS curves in Fig.~\ref{Figure4}(c) exhibit a series of peaks around $V_{\mathrm{D}}$ most likely related to the Sn electronic states in qualitative agreement with the DFT calculations, revealing a complex hybridization scheme of the states close to the Fermi energy.  

Furthermore, we studied the spectra as a function of the set-point current. The results shown in Fig.~\ref{Figure4}(d) reveal a gradual increase in the overall d$I$/d$V$ intensity with increasing current. In particular, the features at $\pm$0.12 eV, attributed to the $S_2$-Sn states, show an enhanced intensity due to the higher probability of tunneling into these states localized beneath the graphene layer. 

To visualize the intercalation of Sn modified electronic structure, we mapped the spatially-resolved DOS distribution along the probed positions, as shown in Fig.~\ref{Figure4}(b). The measured DOS on the bubbles of the Sn-related states is significantly higher than the surrounding regions, indicating a higher coverage of Sn-atoms or smaller distances of the Sn-layer to graphene.

\begin{figure*}[tb]
	\centering
	\includegraphics[width=0.9 \linewidth]{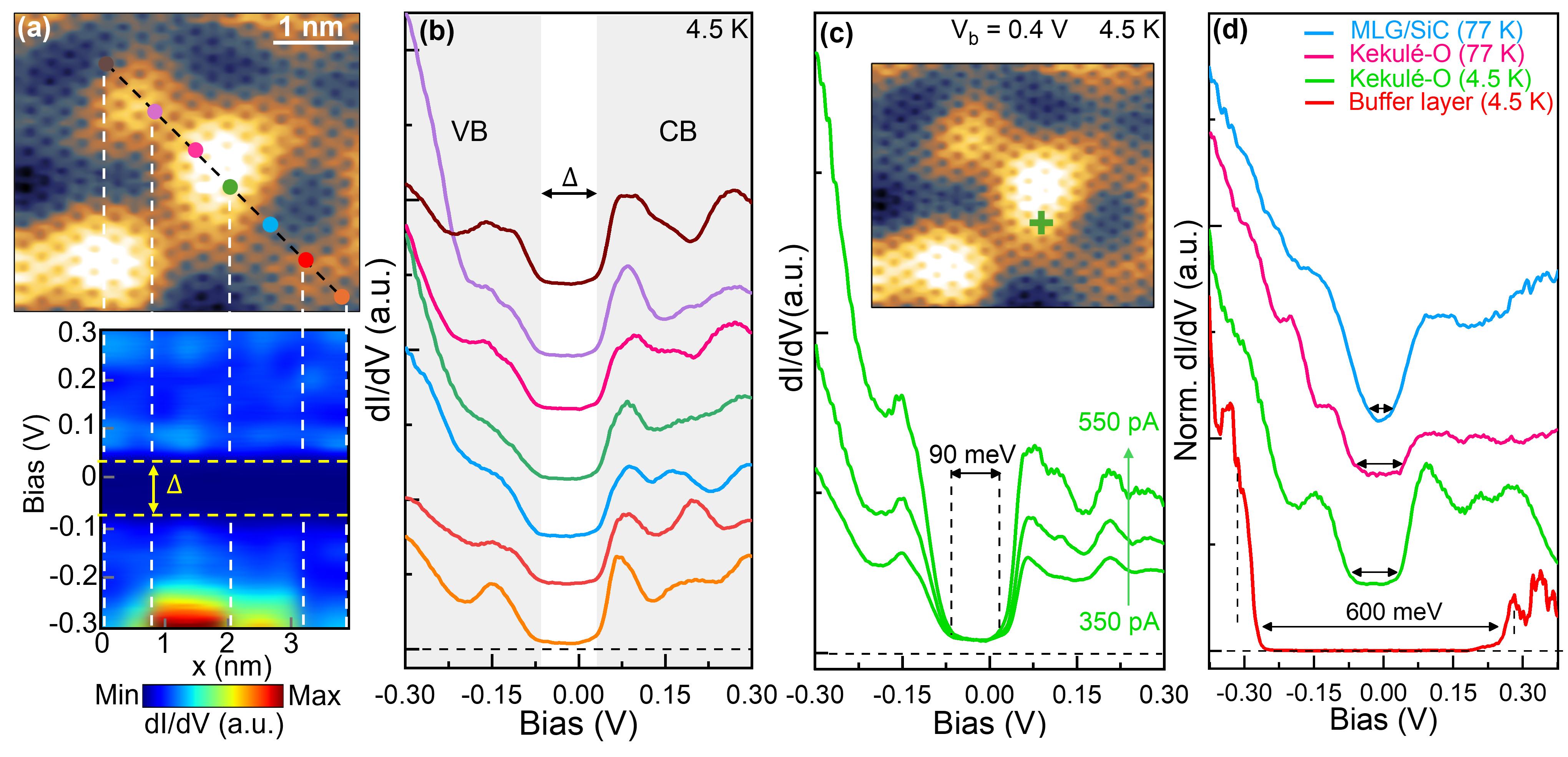}
	\caption{(a) (Top)  STM image of a Sn-intercalated BL/SiC (region II) showing the appearance of Kekul\'e-O ordered graphene at a bias of +0.02 V, as presented in Fig.~\ref{Figure2}(g)-(i). (Bottom) d$I$/d$V$ line scan ($V_b$, x) taken along the black dashed line ($V_b$ = 0.4 V, $I_t$ = 350 pA, T = 4.5 K). (b) The corresponding d$I$/d$V$ spectra extracted from the positions marked by the colored circles in the STM image (Fig.~\ref{Figure5}(a). (c) Current-dependent d$I$/d$V$ spectra of the Kekul\'e-O graphene, exhibiting a band gap of 90 meV ( $V_b$ = 0.4 V, $I_t$ = 350 pA, 450 pA, 550 pA). The inserted STM image shows the probed position. (d) Tunneling d$I$/d$V$ spectra of monolayer graphene (MLG) at 77 K (blue, 0.4 V, 350 pA), buffer layer at 4.5 K (red, 0.4 V, 350 pA), and Kekul\'e-O graphene at 77 K (pink, 0.5 V 300 pA) and at 4.5 K (green, 0.4 V, 350 pA). 
\label{Figure5}} 
\end{figure*}

An elevated local Sn coverage can generate lattice strain in graphene, which in turn facilitates the formation of Kekulé-O patterns \cite{Andrade2019, Mamiyev2025}. Electronically, such an inhomogeneous Sn distribution leads to spatial variations in charge transfer and to a modified electronic coupling between the Sn and graphene layers.  In case that this interaction becomes sufficiently strong, it may promote coupling between the K and K' points in graphene, thereby stabilizing the Kekul\'e-O patterns in graphene. 
In this context, the bubbles seem to act as sites where the graphene–Sn interaction is most intense, providing favorable conditions for modulating the C–C bonds in graphene, e.g., as seen for region II in Fig.~\ref{Figure2}(j) coming along with the opening of an electronic gap (Fig.~\ref{Figure3}(b), green curve). This behavior contrasts sharply with the intercalation of other metals beneath the buffer layer—such as In \cite{Schmitt2024}, Bi \cite{Gehrig@2025}, Pb \cite{Gruschwitz@2025, Ghosal2022}, Pt \cite{Ferbel2025}, and Eu \cite{Anderson@2017}—where Kekul\'e-distorted graphene has not been observed.

To gain deeper insight into the electronic properties, the tunneling spectra shown in Fig.~\ref{Figure5}(b) were measured at the positions indicated in Fig.~\ref{Figure5}(a). The d$I$/d$V$ spectra clearly exhibit an energy gap of approximately 90~meV around the Fermi level. Interestingly, while the background appears inhomogeneous, the band gap in graphene develops uniformly. Although the unoccupied and occupied states around the gap vary slightly between different positions, no in-gap states are observed at any of the probed locations (Fig.~\ref{Figure5}(b)). Likewise, the tunneling spectra display a finite offset in the d$I$/d$V$ intensity, consistent with the conclusions drawn from Fig.~\ref{Figure3}. Furthermore, the d$I$/d$V$ intensities measured directly on the bubble (purple, pink, and green circles) are higher than at other positions, particularly on the negative bias side of the STS map (Fig.~\ref{Figure5}(a), bottom panel). This behavior is consistent with the metallic character of the Sn(1$\times$1) intercalated phase. The minor intensity variations are also clearly visible in the d$I$/d$V$ map shown in Fig.~\ref{Figure5}(a) (bottom panel). 

The band edge features increase in intensity as the set-point current is raised (Fig.~\ref{Figure5}(c)). Moreover, the positions of the VB and CB peaks roughly correlate with the splitting of the $S_2$ state observed in Fig.~\ref{Figure3}(b). In the case of distorted graphene, these positions may vary slightly due to a different hybridization of the $S_2$ state with the graphene bands backfolded to the $\Gamma$-point.

In our case, the observation of the electronic gap is accompanied by a Kekulé distortion in graphene. Nevertheless, several other mechanisms could, in principle, account for the formation of an electronic gap. 
In order to rule out alternatives, we performed measurements on other graphene-like phases. In Fig.~\ref{Figure5}(d), we present a spectrum measured in a region exhibiting the BL structure on our 4H-SiC(0001) substrate. The gap size is around  600 meV, in reasonable agreement with DFT calculations that report 700 meV \cite{Cavallucci2018}.  It should be noted that the BL in this case is formed by Sn deintercalation, and therefore, the structure may not be perfectly ordered as for a pristine BL. 
Furthermore, we measured the tunneling spectrum on MLG, which also exhibits a gap-like feature, an inelastic tunneling gap, reminiscent of that observed in free-standing graphene \cite{Zhang2009}.  In MLG, the gap size of around 125~meV has been attributed to phonon excitation \cite{Zhang@2008}. In contrast, the peak-to-peak separation in our spectra is significantly larger (240~meV), suggesting that a different mechanism is responsible for the observed band gap. One should also bear in mind that inelastic scattering channels, as known for monolayer graphene, play only a secondary role in Kekulé-ordered graphene, since the relevant electronic states are situated near the $\Gamma$ point. 

Moreover, the tunneling spectra of Kekul\'e-O graphene measured at 4.5~K and 77~K exhibit identical electronic features, indicating the temperature insensitivity of the band gap (Fig.~\ref{Figure5}(d), green and pink curves). In particular, the d$I$/d$V$ spectra at both temperatures show a pronounced suppression of the density of states around the Fermi level, demonstrating the Kekul\'e distortion–induced energy gap opening in graphene. At 77~K, the spectral feature appears slightly smoother due to thermal broadening but retains the same gap width, confirming the robust existence of the Kekul\'e-O order over a wide temperature range, even at room temperature as seen in SPA-LEED (cf. Fig.~\ref{Figure1}(g)).  This behavior contrasts with the phonon-induced gap in MLG, which is known to be sensitive to temperature variations \cite{Hacohen-Gourgy@2011}.
Previous studies have shown that the Kekulé O-type sublattice symmetry is associated with an energy gap in graphene of 260 meV on SiC \cite{Zhou2007}, 320 meV on SiO$_x$ \cite{Eom2020}, and 380 meV in Li-intercalated graphene/SiC \cite{Bao2021}.  Although the graphene appears to be freestanding, our  observations suggest that the local Sn($1\times 1$) phases acts as a scattering center. In contrast, such  variations of the surface corrugation were not observed in the case of Pb intercalation.  Unlike Sn, Pb forms long-range ordered mono- and multilayer phases \cite{Gruschwitz@2025, Ghosal2022}. The interaction with graphene appears to be significantly weaker in this case. Nevertheless, a band gap has also been detected in graphene for this system, where the moiré structures induce a symmetry breaking in the graphene on mesoscopic length scales \cite{Ghosal2022}. 

\section{Summary and Conclusion}
We investigated the structural and electronic properties of Sn-intercalated buffer layers on SiC(0001) using LT-STM/STS, SPA-LEED, and DFT calculations. Sn intercalation decouples the buffer layer into quasi–free-standing monolayer graphene while inducing lattice distortions. Bias-dependent STM revealed the coexistence of conventional and Kekul\'e-O graphene domains, governed by Sn(1$\times1$)  at the SiC interface. 
The STS spectra are in very good agreement with the DFT calculations within the accuracy of the underlying model. However, the Sn(1$\times$1) structures cannot be produced homogeneously, and strain within the Sn monolayer appears to be a limiting factor. This is likely also the reason for the formation of Kekul\'e distortions in graphene, which are clearly associated with the opening of a band gap. In the case of much more homogeneous intercalants such as Pb, we did not observe such instabilities. In contrast, adatoms such as Li appear to induce significantly stronger perturbations, as directly reflected by the pronounced band gap in graphene.


\section*{acknowledgments}
The authors thank Christoph Lohse (AG Seyller) for providing zero-layer graphene samples within the DFG Research Unit FOR5242. We gratefully acknowledge financial support from the DFG through projects Te386/22-1 and We5342/7-1. NW acknowledges support from the DFG funded SFB 1170 (``Tocotronics'', project No.~258499086).

%

\end{document}